\documentclass[conference]{IEEEtran}
\IEEEoverridecommandlockouts
\usepackage{cite}

\usepackage{amsmath,amssymb,amsfonts}
\usepackage{algorithmic}
\usepackage{algorithm}
\usepackage{graphicx}
\usepackage{textcomp}
\usepackage{color,soul}
\newcommand{\comment}[1]{}
\usepackage{longtable}
\usepackage{multirow}
\usepackage{color}
\usepackage{caption}
\usepackage{subcaption}
\usepackage{textcomp}
\def\BibTeX{{\rm B\kern-.05em{\sc i\kern-.025em b}\kern-.08em
    T\kern-.1667em\lower.7ex\hbox{E}\kern-.125emX}}
    
\begin{document}

\title{Q-learning-based Joint Design of Adaptive Modulation and Precoding for  Physical Layer Security in Visible Light Communications\\}
\author{\IEEEauthorblockN{Duc M. T. Hoang\IEEEauthorrefmark{1},
Thanh V. Pham\IEEEauthorrefmark{3}, 
Anh T. Pham\IEEEauthorrefmark{4},
and Chuyen T. Nguyen\IEEEauthorrefmark{1}}
\IEEEauthorblockA{\IEEEauthorrefmark{1}Hanoi University of Science and Technology, Vietnam}
\IEEEauthorblockA{\IEEEauthorrefmark{3}Department of Mathematical and Systems Engineering, Shizuoka University, Japan}
\IEEEauthorblockA{\IEEEauthorrefmark{4}Department of Computer Science and Engineering, The University of Aizu, Japan}
}

\maketitle

\begin{abstract}
Physical layer security (PLS) offers a unique approach to protecting information confidentiality against eavesdropping by malicious users. This paper studies a joint design of adaptive $M-$ary pulse amplitude modulation (PAM) and precoding for performance improvement of PLS in visible light communications (VLC). 
It is known that higher-order modulation results in a better secrecy capacity at the expense of a higher bit-error rate (BER). On the other hand, a proper precoding design can also enhance secrecy performance. The proposed design, therefore, aims at the optimal PAM modulation order and precoder to maximize a utility function that takes into account the secrecy capacity and BERs of the legitimate user (Bob)'s and the eavesdropper (Eve)'s channel. Due to the lack of a closed-form expression for the utility function, a Q-learning-based design is proposed and evaluated. Compared to the non-adaptive approach under all different settings of Bob's and Eve's positions, simulation results verify that the proposed joint adaptive design achieves a good balance between the secrecy capacity and BER of Bob's channel while maintaining a sufficiently high BER of Eve's channel.  
\end{abstract}


\section{Introduction}

In recent years, visible light communications (VLC) has received considerable attention from the research community as a promising wireless technology to complement existing systems \cite{arfaoui2020physical}. Compared with the radio frequency (RF) counterparts, VLC can take advantage of the license-free visible light spectrum and the widespread deployment of LED in the lighting infrastructure. 

As an emerging technology for the next wireless generation, there have been extensive theoretical and experimental studies on the performance and practicality of VLC systems. In addition to these aspects, ensuring information confidentiality is another indispensable requirement. This is even more concerning in the case of wireless systems due to the broadcast nature of the signal. 
In this regard, physical layer security (PLS)  has emerged as a novel technique that can provide unconditional security by exploiting the physical properties of the communication channel \cite{wyner1975wire,csiszar1978broadcast}. It is in sharp contrast with the computational security (a.k.a conditional security) provided by traditional key-based cryptographic techniques, which are considered vulnerable to the computing capability of quantum computers \cite{Davide2022}. The performance of PLS is characterized by the secrecy capacity, which measures the maximum rate between the sender and the receiver at which no information is leaked to the eavesdropper regardless of its computational power.

For indoor scenarios, multiple LED luminaires are often deployed to provide sufficient illumination. As a result, multiple-input single-output (MISO) wiretap channels are of practical interest. The spatial degrees of freedom brought by multiple transmitters enables the use of precoding to enhance the PLS performance. In this context, a great deal of research effort has been dedicated to studying the different precoding schemes under the signal amplitude constraint of VLC systems \cite{mostafa2015physical,Ma2016, Pham2017, Arfaoui2018}. To facilitate solving the design optimization problems, simple closed-form expressions for the lower and upper bounds of the secrecy capacity were utilized to formulate the objective functions. Note that these bounds were derived independently from the modulation scheme. As a result, evaluations of signal quality parameters such as the bit error rate (BER) were ignored. In summary, while the derived bounds are useful in understanding the secrecy performance limits, they offer little insight into the performance from a practical design with specific modulations.  

Motivated by these observations, this paper attempts a joint design of adaptive $M$-ary pulse amplitude modulation (PAM) and precoding that takes into account secrecy capacity and BER performance. For this purpose, we first define a utility function that includes the secrecy capacity and BERs of the legitimate user and the eavesdropper. Due to the lack of closed-form expressions for the PAM-constrained secrecy capacity, solving optimization problems that maximize the utility function is challenging (if not impossible). Therefore, we present in this paper a Q-learning-based transmission scheme that adaptively selects the optimal modulation order and precoder. Under three different settings of Bob's and Eve's positions, simulation results show that  the joint design can adaptively select the optimal modulation order and precoder to achieve higher utility values compared to the non-adaptive design, which indicates a better balance between the secrecy capacity and BER of Bob's channel. In addition, Eve's BERs are controlled to be sufficiently high (i.e., higher than 0.2), thus ensuring a low information leakage.

The paper is structured as follows. In Section II, the channel and signal models of the system are described. The secrecy capacity and BERs of for $M-$ary PAM are given in Section III. Section IV presents the Q-learning-based algorithm for the proposed joint adaptive modulation and precoding. Simulation results and discussions are given in Section V. Finally, Section VI concludes the paper. 

Notation: The following notations are used throughout the paper. Lowercase bold letters (e.g., $\mathbf{x}$) represent vectors. $I(\cdot;\cdot)$, $h(\cdot)$, and $p(\cdot)$ denote the mutual information, differential entropy, and probability of an event.  
\section{System Model} \label{Sect03}
\subsection{Channel Model}
\label{Section_Channel}

\begin{figure}
    \centering
    \includegraphics[width = .45\textwidth]{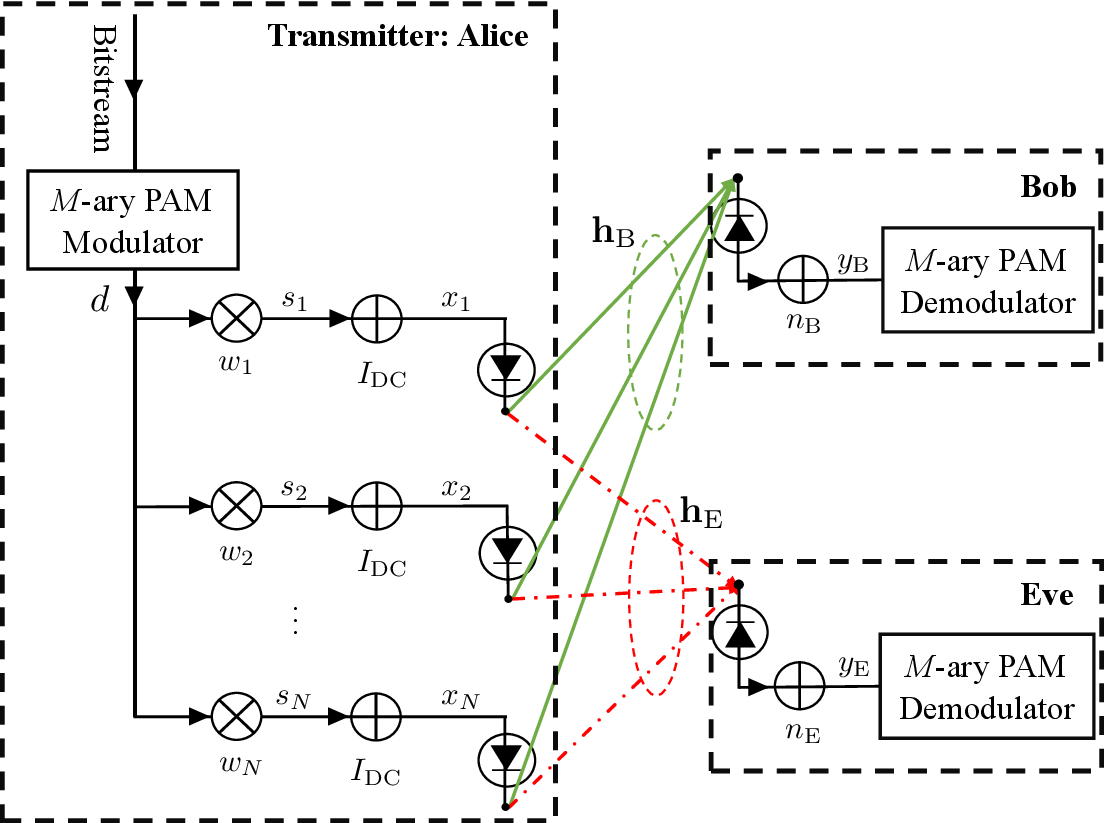}
    \caption{System model.}
    \label{fig:1}
\end{figure}

As shown in Fig.~\ref{fig:1}, we consider in this paper a VLC system with $N$ LED luminaires, a legitimate user (Bob), and an eavesdropper (Eve). Both Bob and Eve are equipped with a single-photodiode (PD) receiver. It is assumed that Eve is an ordinary user of the network who, however, also tries to wiretap Bob's channel. The presence of Eve is thus known by the transmitter. 

In indoor environments, VLC channels comprise two components, namely line-of-sight (LoS) and non-line-of-sight (NLoS) due to reflections of walls and ceilings. It is shown in \cite{komine2004fundamental} that the LoS propagation accounts for more than 95\% of the total received optical power at the receiver. Hence, for the sake of simplicity, it is reasonable to only take the LoS channel into consideration. Denote $\mathbf{h}_{\text{R}}= \begin{bmatrix}h^1_{\text{R}} & h^2_{\text{R}} & \cdots & h^N_{\text{R}}\end{bmatrix}^T \in \mathbb{R}^N_{\geq 0}$ as the channel vector whose $n$-th element $h^n_{\text{R}}$ represents the  gain of the channel between the $n$-th luminaire and the receiver\footnote{The subscript `R' is used to denote Bob and Eve in general. It is replaced by `B' and `E' to explicitly refer to Bob and Eve, respectively, when necessary.}. According to \cite{komine2004fundamental}, $h^n_{\text{R}}$ is given by
\begin{align}
    \label{eq:2}
   h^n_{\text{R}} = \left\{ \begin{array}{l}
\frac{{{A_r}}}{{d_n^2}}L\left( {{\phi}} \right){T_s}\left(\psi \right)g\left( \psi\right)\cos \left( \psi \right), \hspace{2mm} 0 \le {\psi} \le {\Psi}, \\
0                           , \hspace{2mm}{\psi} > {\Psi},
\end{array} \right.
\end{align}
where $A_r$, $d_n$, $\phi$, $\psi$, $\Psi$ are the active area of the PD, the distance between the luminaire and the PD, the LED's angle of irradiance, the angle of incidence, and the optical field of view (FoV) of the PD, respectively. Assuming the Lambertian beam distribution gives the emission intensity $L(\phi) = \frac{l+1}{2\pi}\cos^l(\phi)$ where the Lambertian emission order is given by
\begin{align}
    l = \frac{\ln(2)}{\ln(\Theta_{0.5})},
\end{align}
where $\Theta_{0.5}$ is the LED’s semi-angle for half illuminance. In addition, $T_s(\psi)$ denotes the gain of the optical filter $g(\psi)$ is the gain of the optical concentrator, which is given by
\begin{align}
   g(\psi) =  \left\{ \begin{array}{l}
    \frac{\kappa^2}{\sin^2(\Psi)}, \hspace{2mm} 0 \le {\psi} \le {\Psi}, \\
    0, \hspace{2mm}{\psi} > {\Psi},
    \end{array} \right.
\end{align}
where $\kappa$ is the refractive index of the optical concentrator. 

In this paper, Eve is assumed to be an ordinary user who tries to eavesdrop on Bob's channel. Eve's channel vector $\mathbf{h}_{\text{E}}$ can be known at Alice.  
\subsection{Signal Model} 
$M$-ary PAM modulation is widely employed for VLC systems due to its implementation simplicity. At the transmitter, the generated bitstream is modulated to a PAM signal $d$. At the $n$-th luminary, the signal is first precoded with a weight $w_n$ and then is 
added a DC-bias current $I_{\text{DC}}$. The drive current of the LEDs is thus given by
\begin{align}
    x_n = w_nd+ I_{\text{DC}}.
\end{align}
To guarantee LEDs to work within their linear range, $x_n$ should be constrained within $I_{\text{DC}} \pm \alpha I_{\text{DC}}$ where $\alpha \in [0, ~1]$ is the modulation index. 
This can be satisfied by constrainting $|w_n| \leq 1$ and $|d| \leq \alpha I_{\text{DC}}$. 

For the sake of conciseness, the transmit signal vector can be written as
\begin{align}
    \mathbf{x} = \mathbf{w}d + \mathbf{1}_NI_{\text{DC}}, 
\end{align}
where $\mathbf{x} = \begin{bmatrix}x_1 & x_2 & \cdots & x_N\end{bmatrix}^T \in \mathbb{R}^N$ and $\mathbf{w} = \begin{bmatrix}w_1 & w_2 & \cdots & w_N\end{bmatrix}^T \in \mathbb{R}^N$ is the transmit precoder. 
Denoting $\eta$ as the conversion efficiency of the LEDs, the emitted optical power vector is given by
\begin{align}
    \mathbf{p}_t = \eta \mathbf{x} = \eta(\mathbf{w}d + \mathbf{1}_NI_{\text{DC}}).
\end{align}
The electrical signal received by the user is then written by
 \begin{equation}
 \begin{aligned} \label{eq:3}
 y_{\text{R}} &= \gamma \mathbf{h}^T_{\text{R}}\mathbf{p}_t + n_{\text{R}} \\
 &  =\gamma \eta \mathbf{h}^T_{\text{R}}\left(\textbf{w}d + \mathbf{1}_NI_{\text{DC}}\right) + n_{\text{R}},
\end{aligned}
 \end{equation}
 where $\gamma$ is the PD responsivity and $n_{\text{R}}$ is assumed to be zero-mean additive white Gaussian noise with variance $\sigma^2_{\text{R}}$. For signal modulation, the DC-term $\mathbf{h}^T_{\text{R}}\mathbf{1}_NI_{\text{DC}}$, which contains no information, is filtered out, resulting in 
\begin{align}
    \overline{y}_{\text{R}} =\gamma \eta \mathbf{h}^T_{\text{R}}\textbf{w}d  + n_{\text{R}}.
    \label{received_demodulation_signal}
\end{align}
The constraint on the precoder $\mathbf{w}$ can also be represented as
\begin{align}
    \left\lVert\mathbf{w}\right\rVert_{\infty} \leq 1,
    \label{precoder-constraint}
\end{align}
where $\lVert\cdot\rVert_{\infty}$ denotes the infinity norm. 
\section{Secrecy Capacity and Bit Error Rate}
In this section, the secrecy capacity  of the wiretap channel described in \eqref{received_demodulation_signal} and the BERs at Bob and Eve are derived. 
\subsection{Secrecy Capactiy}
The secrecy capacity $C_s$ of the wiretap channel characterized by \eqref{received_demodulation_signal} is given by \cite{csiszar1978broadcast}
\begin{align} 
    C_s= {I(d;\overline{y}_{\text{B}}) - I\left( {d;\overline{y}_{\text{E}}} \right)},
    \label{secrecy-rate}
\end{align}
where the two mutual information are given by
    
 \begin{align}
 \label{eq:7}
&I(d;\overline{y}_{\text{R}}) = h(\overline{y}_{\text{R}}) - h(\overline{y}_{\text{R}}|d)=h(\overline{y}_{\text{R}}) - h(n_\text{B}),
 \end{align}
The differential entropies of $\overline{y}_{\text{R}}$ and $n_{\text{R}}$ are written by 
\begin{align}\label{eq:8}
&h({\overline{y}_{\text{R}}}) = \int _{-\infty}^{+\infty}{p({\overline{y}_{\text{R}}})} {\log _2}p({\overline{y}_{\text{R}}})\text{d}{\overline{y}_{\text{R}}},\\
&h({n_\text{R}}) = {\log _2}\pi e\sigma^2_{\text{R}},
\end{align}
respectively. Let $\{d_i\}$ with $i = 0, 1, \cdots, M-1$ be the constellation points and assume that these are transmitted with equal probabilities (i.e., $p(d_i) = \frac{1}{M}, ~\forall i = 0, 1, \cdots, M-1$). Since $n_{\text{R}} \sim \mathcal{N}(0, \sigma^2_{\text{R}})$,  ${p({\overline{y}_{\text{R}}})}$ is given by
\begin{align}\label{eq:9}
p({\overline{y}_{\text{R}}}) = & \sum_{i = 0}^{M - 1}p(\overline{y}_{\text{R}}|d = d_i)p(d_i) \nonumber \\ = &
\frac{1}{M}\sum\limits_{i = 0}^{M - 1} {\frac{1}{\sqrt{2\pi}\sigma_{\text{R}} }} \exp \left({ - \frac{{{\left| {{\overline{y}_{\text{R}}} - \gamma\eta\textbf{h}^T_\text{R} \textbf{w}{d_i}} \right|}^2}}{2\sigma^2_{\text{R}}}} \right).
\end{align}
\subsection{Bit Error Rate}
According to \cite[Eqs.~(9) and (10)]{Kyongkuk2002}, the BER at Bob and Eve can be calculated as
    
 \begin{align}
{p_\text {e,R}} = &\sum\limits_{k = 1}^{{{\log }_2}M} {\sum\limits_{i = 0}^{M(1 - {2^{ - k}}) - 1}\!\!\!\!\!\!{{{( - 1)}^{\left\lfloor {\frac{{i{{2}^{k - 1}}}}{M}} \right\rfloor }}} }  \left( {{2^{k - 1}} - \left\lfloor {\frac{{i2^{k - 1}}}{M} + \frac{1}{2}} \right\rfloor } \right) \nonumber \\ & \times \frac{1}{{M{{\log }_2}M}}  \text{erfc}\left( (2i + 1)
\sqrt{\frac{3\left(\gamma\eta\mathbf{h}^T_{\text{R}}\mathbf{w}/\sigma_{\text{R}}\right)^2E_s}{M^2-1}} \right),
\label{ber}
\end{align}
 where $\lfloor\cdot\rfloor$ is the floor function, and $\text{erfc}(x) = \frac{2}{\sqrt{\pi}}\int^{\infty}_xe^{-t^2}\text{d}t$ is the complimentary error function. $E_s$ is the average symbol energy.

\section{Q-Learning-based Joint Design of Adaptive PAM Modulation and Precoding}\label{Sect04}

In this study, the purpose of the joint adaptive modulation and precoding design is to achieve a high secrecy capacity while attaining a low BER at Bob and a high BER at Eve. For this, a utility function, which takes into account these parameters, is defined as follows 
\begin{align}
    u = C_s - \delta p_{e,\text{B}} + \zeta p_{e,\text{E}},
    \label{utility-function}
\end{align}
where $\delta$ and $\zeta$ are chosen coefficients that balance the contribution of Bob and Eve's BER and secrecy capacity to the utility function. From the system perspective, $\delta$ and $\zeta$  set a trade-off between the secrecy capacity and communication quality of Bob's and Eve's channels. Ideally, one should formulate an optimization problem that maximizes $u$ under the constraint given in \eqref{precoder-constraint} such as
\begin{subequations}
\begin{alignat}{2}
\bf{\mathcal{P}1} \hspace{5mm} & \underset{M, \mathbf{w}}{\text{maximize}}  & \hspace{2mm}  & u \label{P1:obj_func}\\
& \text{subject to} &		& \nonumber \\
&  & & \lVert\mathbf{w}\rVert_{\infty} \leq 1. \label{P1:constraint2} 
\end{alignat}
\end{subequations}
However, the lack of a simple closed-form expression for u (due to the integral form of $C_s$) would render solving $\bf{\mathcal{P}1}$ challenging (if not impossible). 

The difficulty of a rigorous mathematical solution to the above optimization problem motivates us to investigate a Q-learning-based transmission control scheme that maximizes $u$ through a trial-and-error learning process. Firstly, let $\textbf{s}$ be a state, which contains the previous BERs of Bob and Eve obtained from the previous state and the current channel vectors of Bob and Eve. Let $\mathbf{a}$ be an action to be taken, which consists of a modulation order $m$ and a precoder $\mathbf{w}$. Define $\mathbb{A}$ as the set of all possible actions. Note that the size of $\mathbb{A}$ sets a trade-off between the achievable performance and the computational complexity. For the action space of the modulation order, 2, 4, 8, 16, 32, and 64-PAM constellations are considered. For the precoder, note that each element $w_i~(i = 1, 2, \cdots, N)$ of $\mathbf{w}$ is constrained between $[-1,~1]$. To form a finite-size action space of the precoder,  $w_i$ is quantized into $2T_s+1$ discrete equally spaced values, i.e., $w_i \in \left\{\frac{t}{T_s}~\big| -T_s \leq t \leq T_s,~t \in \mathbb{Z}\right\}$ \cite{Xiao2019}. 


Q-learning achieves an optimal modulation order and precoder through a sufficient long learning iteration. Specifically, at  the time slot $k$, the transmitter Alice selects an action $\textbf{a}^{(k)} = \begin{bmatrix}m^{(k)}, \mathbf{w}^{(k)}\end{bmatrix}$ based on the current state $\textbf{s}^{(k)}$, which consists of the BERs of Bob and Eve, the secrecy capacity at the previous time slot and the channel vectors of Bob and Eve at the current time slot, i.e., $\textbf{s}^{(k)} = \begin{bmatrix}p^{(k-1)}_{e,\text{B}},p^{(k-1)}_{e,\text{E}}, C^{(k-1)}_s, \mathbf{h}^{(k)}_{\text{B}}, \mathbf{h}^{(k)}_{\text{E}}\end{bmatrix}$. After applying the action, Alice calculates the current secrecy rate and BERs according to \eqref{secrecy-rate}-\eqref{ber} and evaluates the current utility function $u^{(k)}$ using \eqref{utility-function}.
where $\delta$ and $\zeta$ is the coefficient to balance the contribution of Bob and Eve's BER and secrecy capacity to the utility function.  
With the observation of the environment through the utility function, Alice constructs the Q-value table and updates it using the Bellman equation as follows \cite{Mnih2015}
\begin{align}
    Q\left({\textbf{s}^{(k)}}, \textbf{a}^{(k)}\right) & =  (1 - \lambda )Q\left({\textbf{s}^{(k)}},{\textbf{a}^{(k)}}\right) \nonumber \\ 
              & + \lambda \left({u^{(k)}} + \beta {\max _{\textbf{a}}}~Q\left({\textbf{s}^{(k + 1)}},{\textbf{a}^{(k + 1)}
    }\right)\right).
    \label{BellManEq}
\end{align}
Here, $\lambda \in [0,1]$ is the learning rate, which determines the significance of the current Q-value, while $\beta \in [0,1]$ is the discount parameter, which represents the weight given to long-term rewards. The updated Q-value represents the immediate reward of a policy and the discounted reward of the previous action in the given state. 

To optimize selecting the action, it is assumed that Alice employs the $\epsilon$-greedy algorithm to balance between exploration and exploitation, which is described as
\begin{equation}
P\left({{\rm{\textbf{a}}}^{(k)}} = {\rm{\textbf{a}^*}}\right) = \left\{ \begin{array}{l}
1 - \varepsilon ,~~{{ \textbf{a}^* = }}\arg \mathop {\max }\limits_{{\rm{\textbf{a}}}'} Q\left({\textbf{s}^{(k)}},{\rm{\textbf{a}}}'\right)\\
\varepsilon , ~~~~~~~\text{otherwise}.
\end{array} \right.
\end{equation}
The design scheme initially takes random actions with a probability of $\epsilon$ in order to explore the environment and avoid getting stuck in local optima. As the number of time slots increases and $\epsilon$ decreases, the design increasingly favors selecting the action with the maximum Q value with probability of $1-\epsilon$, exploiting the highest existing Q value, and optimizing the modulation order and precoder. A summary of the Q-learning-based joint design scheme is given below.
\begin{algorithm}
\caption{Q-learning-based Joint Design of Adaptive Modulation and Precoding}\label{alg:cap}
\begin{algorithmic}[1]
\STATE $Q(\textbf{s}, \textbf{a}) = 0,~\forall \textbf{s} \in \Lambda,~\textbf{a} \in \mathbb{A}$ 
\STATE Initialize $\textbf{a}^{(0)}= [m^{(0)},~\textbf{w}^{(0)}]$, discount factor $\beta$ and learning rate $\lambda$
\FOR{$k=1,2,3,...$}
\STATE Calculate the BERs and secrecy capacity from the previous time slot $p_\text{e,B}^{(k-1)}$, $p_\text{e,E}^{(k-1)}$, and $C_s^{(k-1)}$
\STATE Form the current state ${\textbf{s}^{(k)}}$, which is defined as ${\textbf{s}^{(k)}} = [p_\text{e,B}^{(k - 1)}, p_\text{e,E}^{(k - 1)}, C_\text{s}^{(k - 1)}, \mathbf{h}^{(k)}_{\text{B}}, \mathbf{h}^{(k)}_{\text{E}}]$
\STATE Choose an action $\textbf{a}^{(k)} = [m^{(k)}, ~ \mathbf{w}^{(k)}]$ using the $\epsilon$-greedy method
\STATE Transmit the signal with the action $\textbf{a}^{(k)}$
\STATE Calculate the current BERs $p_\text{e,B}^{(k)}$,  $p_\text{e,E}^{(k)}$ and the secrecy capacity $C_s^{(k)}$
\STATE Calculate the utility and update the Q-value according to \eqref{BellManEq}.
\ENDFOR
\end{algorithmic}
\end{algorithm}

\section{Simulation Results and discussions}
In this section, simulation results are provided to highlight the performance of the proposed joint design. Without otherwise noted, simulation parameters are given in Table I. Note that, the positions of LED luminaries, Bob, and Eve are specified with respect to a 3D Cartesian coordinate system whose center is the floor center. The learning process is performed over 2000 time slots, during which the first 600 time slots of the learning process, the $\epsilon$-greedy parameter is gradually reduced from 1.0 to 0.1 to encourage exploration. After this initial phase, the parameter is set to 0.1 to ensure stability and exploit the best current action.

To illustrate the adaptivity of the proposed scheme to different channel conditions, three different positions of Bob and Eve given in Table II are considered for simulations. 
\begin{table}[H]
    \centering
    \begin{tabular}{l l}
\hline
     \textbf{Parameter} & \textbf{Value}\\\hline
     Room Dimension     & 10 m $\times$ 10 m $\times$ 3 m  \\\hline
      Positions of LED luminaires   & $\begin{bmatrix}
      -\sqrt{5}, -\sqrt{5}, 3\end{bmatrix}$ $\begin{bmatrix}\sqrt{5}, -\sqrt{5}, 3\end{bmatrix}$\\ &$\begin{bmatrix}\sqrt{5}, \sqrt{5}, 3\end{bmatrix}$~~~~~$\begin{bmatrix}-\sqrt{5}, \sqrt{5}, 3\end{bmatrix}$\\ \hline
      Transmit optical power per luminaire, $p_t$ & 5W \\ \hline
      LED beam angle, $\phi$    & $120^\circ$ \\\hline
      LED conversion factor, $\eta$     & 0.44 W/A  \\\hline
      PD active area, $A_r$     & 1 $\text{cm}^2$  \\\hline
      PD responsivity, $\gamma$     & 0.54 A/W  \\\hline
      PD field of view (FoV), $\Psi$     & $120^\circ $  \\\hline
     Optical filter gain, $T_s(\psi)$      & 1  \\\hline
     Refractive index of concentrator, $\kappa$ &1.5  \\\hline
     Modulation index, $\alpha$      & 0.1  \\\hline
     Average noise power, $\sigma_{\text{R}} ^2$      & -98.82 dBm  \\\hline
 Bob's BER coefficient, $\delta$      & 10  \\\hline
 Eve's BER coefficient, $\zeta$ & 5  \\\hline
       Learning rate, $\lambda$      & 0.5  \\\hline
    Discount parameter, $\beta$ & 0.5  \\\hline
    \end{tabular}
    \caption{System Parameters}
    \label{tab:System parameters}
\end{table}
\begin{table}[htb]
    \centering
    \resizebox{0.35\textwidth}{!}{\begin{tabular}{|c|c|c|}
\hline
     & \textbf{Bob's position }&\textbf{Eve's position}\\\hline
     \textbf{Setup 1}     & $\begin{bmatrix}0, 0, 0.5\end{bmatrix}$  & $\begin{bmatrix}1, 0, 0.5\end{bmatrix}$\\\hline
      \textbf{Setup 2}    & $\begin{bmatrix}-1, -1, 0.5\end{bmatrix}$ & $\begin{bmatrix}-3, -3, 0.5\end{bmatrix}$\\\hline
     \textbf{Setup 3}  & $\begin{bmatrix}-2, -2, 0.5\end{bmatrix}$ &$\begin{bmatrix}-5, -5, 0.5\end{bmatrix}$\\ \hline
    \end{tabular}
    }
    \caption{Setup's positions.}
    \label{tab:my_label}
\end{table}

To illustrate the adaptivity of the proposed design to different channel conditions, three different positions of Bob and Eve given in Table II are examined. Also, the performance of the proposed joint adaptive design is compared with that of the non-adaptive one whose modulation order is fixed to 64. 
\begin{figure}[ht]
    \centering
    \includegraphics[width = 0.5\textwidth,height = 6.2cm]{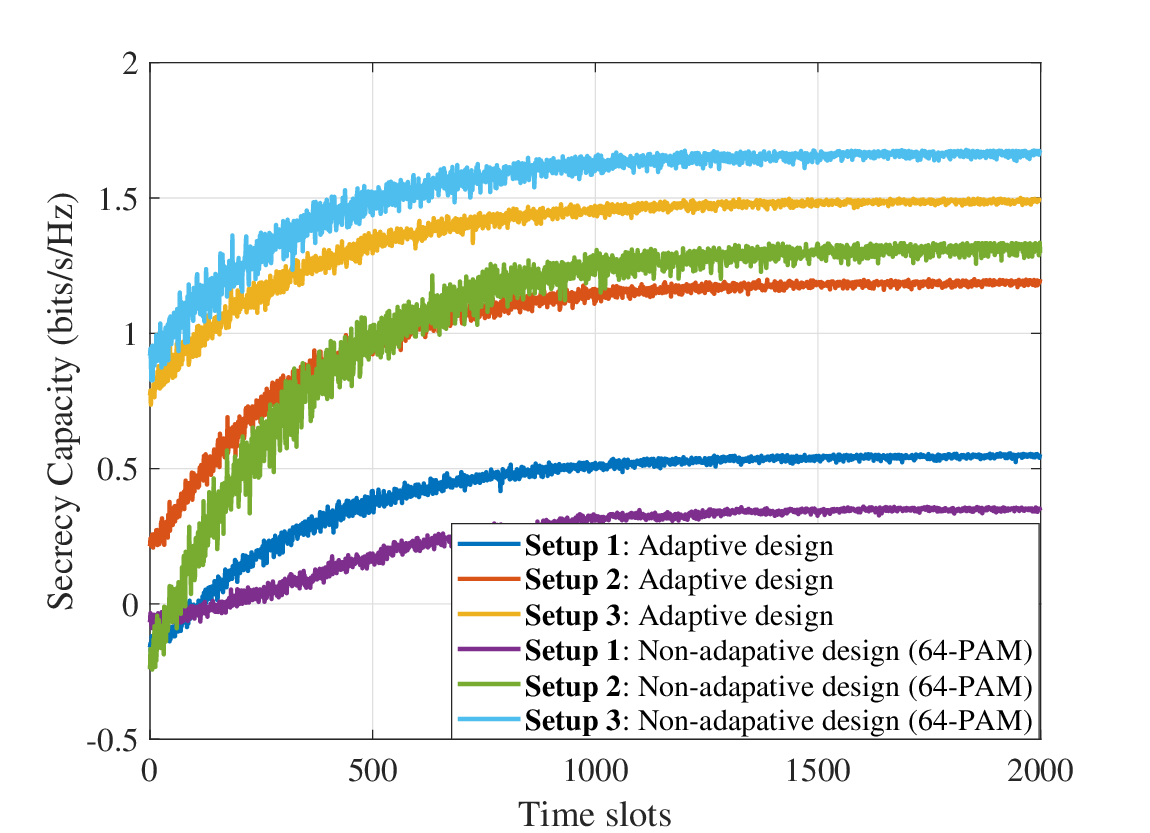}
    \caption{Comparision of the secrecy capacity.}
    \label{fig:secrecycapacity}
\end{figure}

\begin{figure}[ht]
    \centering
    \includegraphics[width = 0.5\textwidth,height = 6.2cm]{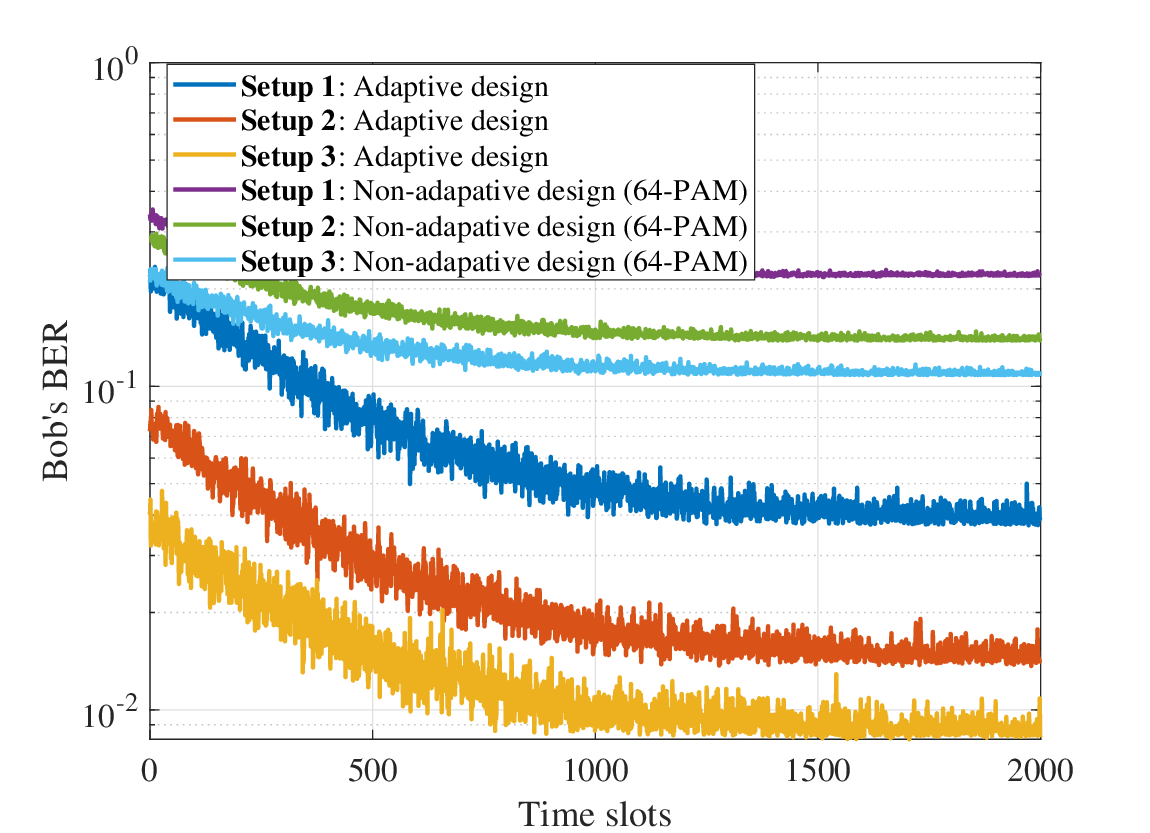}
    \caption{Comparision of Bob's BERs. }
    \label{fig:BobBER}
\end{figure}
\begin{figure}[ht]
    \centering
    \includegraphics[width = 0.5\textwidth,height = 6.2cm]{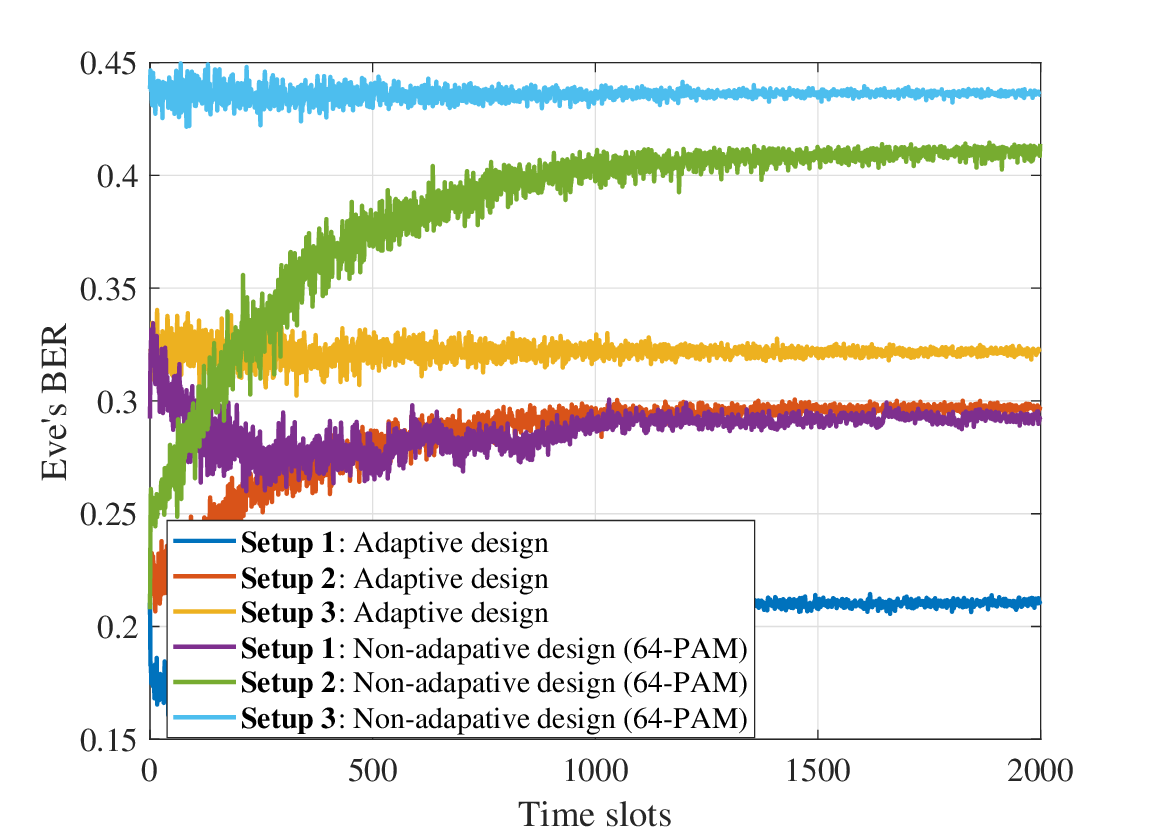}
    \caption{Comparision of Eve's BERs.}
    \label{fig:EveBER}
\end{figure}
\begin{figure}[ht]
    \centering
    \includegraphics[width = 0.5\textwidth,height = 6.2cm]{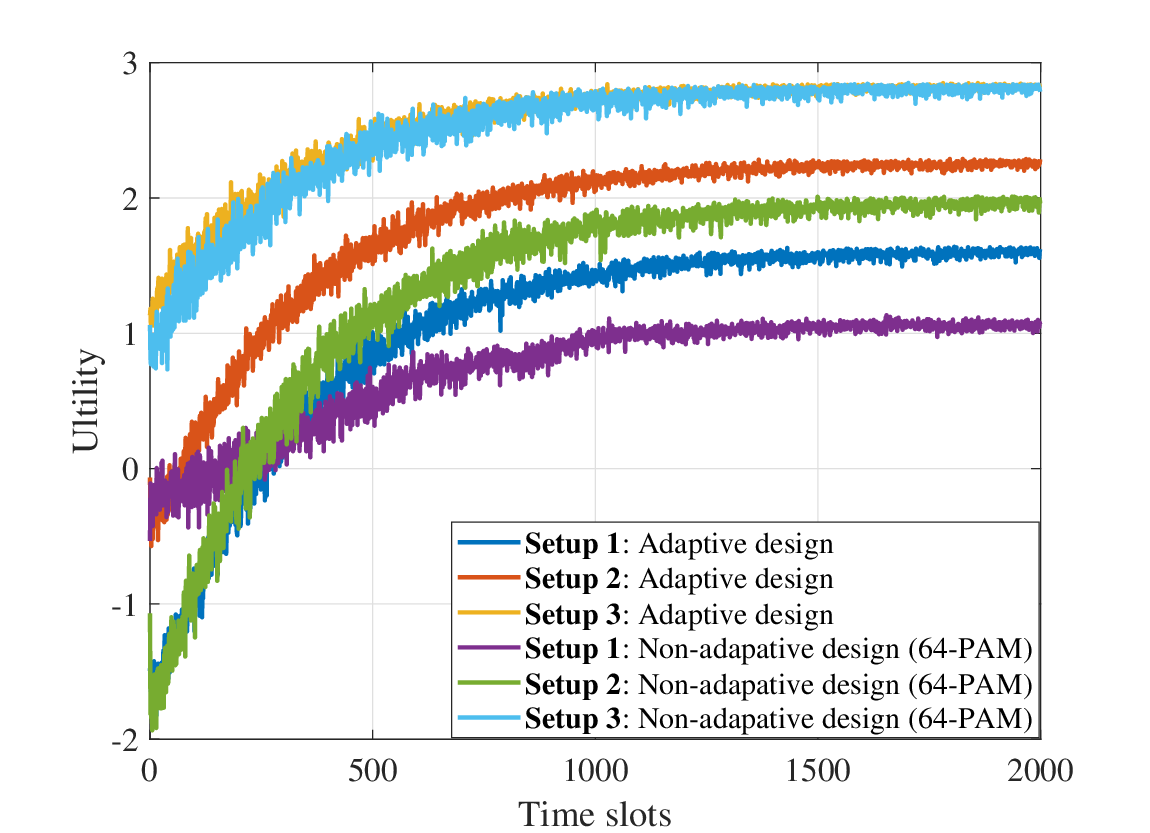}
    \caption{Comparision of utility.}
    \label{fig:Utility}
\end{figure}

Firstly, Figs.~\ref{fig:secrecycapacity}, ~\ref{fig:BobBER}, and \ref{fig:EveBER} show comparisons of the secrecy capacity, Bob's and Eve's BER among the three settings, respectively. It is observed that the secrecy capacity of the joint adaptive design in $\textbf{Setup 1}$ is better than that of the non-adaptive one by about 0.25 bits/s/Hz. On the other hand, in $\textbf{Setup 2}$ and $\textbf{Setup 3}$, the nonadaptive design achieves 0.15 and 0.2 bits/s/Hz higher secrecy capacities, however, at the expense of very high BER in Bob (i.e., about $2\times10^{-1}$ and $10^{-1}$ compared with $3\times 10^{-2}$ and $10^{-2}$). It is also seen that both the adaptive and non-adaptive designs can be able to control Eve's BERs sufficiently high (i.e., higher than 0.2), thus reducing the probability of information leakage. Overall, as depicted in Fig.~\ref{fig:Utility}, the adaptive design maintains better utilities in $\textbf{Setup 1}$ and  $\textbf{Setup 2}$  while achieving the same performance as the non-adaptive one in $\textbf{Setup 3}$. 

\section{Conclusion}\label{Sect05}
This paper has studied a Q-learning-based joint design of adaptive $M$-ary PAM and precoding to improve the performance of PLS in VLC systems. The design takes into account the secrecy capacity and the BERs at Bob and Eve. Simulation results revealed that the joint adaptive design can strike a good balance between the secrecy capacity and BER of Bob's channel while maintaining a sufficiently high BER of Eve's channel, which is critical in guaranteeing a low probability of being eavesdropped by Eve.  


\bibliographystyle{ieeetr}
\bibliography{ref_qlearningVLC}
\vspace{12pt}

\end{document}